\documentclass[10pt,twocolumn,letterpaper]{article}

\usepackage[review,algorithms]{wacv}      % To produce the REVIEW version for the algorithms track

\usepackage[T1]{fontenc}
\usepackage{amsmath}
\usepackage{multirow}
\usepackage{pbox}
\usepackage{utfsym}
\usepackage[shortlabels]{enumitem}
\usepackage{amssymb}

\usepackage{graphicx}
\usepackage{amssymb}
\usepackage{booktabs}
\usepackage{enumitem}
\usepackage{amssymb}
\usepackage{amsmath}
\usepackage{bm}
\usepackage{multirow}
\usepackage{pbox}
\usepackage{graphicx}
\usepackage{xfakebold}
\newcommand{\fbseries}{\unskip\setBold\aftergroup\unsetBold\aftergroup\ignorespaces}
\makeatletter
\newcommand{\setBoldness}[1]{\def\fake@bold{#1}}
\makeatother
\usepackage[pagebackref,breaklinks,colorlinks]{hyperref}

\usepackage[capitalize]{cleveref}
\crefname{section}{Sec.}{Secs.}
\Crefname{section}{Section}{Sections}
\Crefname{table}{Table}{Tables}
\crefname{table}{Tab.}{Tabs.}

\def\confName{WACV}
\def\confYear{2025}

\newcommand\figref{Figure~\ref}
\newcommand\tabref{Table~\ref}

\usepackage[letterpaper,top=1in,bottom=1in,left=0.75in,right=0.75in]{geometry}

\begin{document}

\title{\textnormal{This paper has been accepted for WACV 2025} \\[1ex]
Compositional Segmentation of Cardiac Images Leveraging Metadata}

\author{
    Abbas Khan\\
    School of Electronic Engineering and Computer Science, Queen Mary University of London, UK\\
    Queen Mary's Digital Environment Research Institute (DERI), London, UK\\
    {\tt\small acw676@qmul.ac.uk}
    \and
    Muhammad Asad\\
    School of Biomedical Engineering and Imaging Sciences King’s College London, UK\\
    Queen Mary's Digital Environment Research Institute (DERI), London, UK\\
    {\tt\small muhammad.asad@qmul.ac.uk}
    \and
     Martin Benning\\
     Department of Computer Science, University College London, UK \\
    Queen Mary's Digital Environment Research Institute (DERI), London, UK\\
    {\tt\small martin.benning@ucl.ac.uk}
    \and
    Caroline Roney\\
    School of Engineering and Materials Science, Queen Mary University of London, UK\\
    Queen Mary's Digital Environment Research Institute (DERI), London, UK\\
    {\tt\small c.roney@qmul.ac.uk}
    \and
    Gregory Slabaugh\\
    School of Electronic Engineering and Computer Science, Queen Mary University of London, UK\\
    Queen Mary's Digital Environment Research Institute (DERI), London, UK\\
    {\tt\small g.slabaugh@qmul.ac.uk}
}

\maketitle
%%%%%%%%% ABSTRACT
\begin{abstract}
   Cardiac image segmentation is essential for automated cardiac function assessment and monitoring of changes in cardiac structures over time. Inspired by coarse-to-fine approaches in image analysis, we propose a novel multi-task compositional segmentation approach that can simultaneously localize the heart in a cardiac image and perform part-based segmentation of different regions of interest. We demonstrate that this compositional approach achieves better results than direct segmentation of the anatomies. Further, we propose a novel Cross-Modal Feature Integration (CMFI) module to leverage the metadata related to cardiac imaging collected during image acquisition. We perform experiments on two different modalities, MRI and ultrasound, using public datasets, Multi-Disease, Multi-View, and Multi-Centre (M\&Ms-2) and Multi-structure Ultrasound Segmentation (CAMUS) data, to showcase the efficiency of the proposed compositional segmentation method and Cross-Modal Feature Integration module incorporating metadata within the proposed compositional segmentation network. The source code is available: https://github.com/kabbas570/CompSeg-MetaData.
\end{abstract}
%%%%%%%%% BODY TEXT
\section{Introduction}
\label{sec:intro}
Segmenting cardiovascular anatomies in cardiac imaging involves dividing the image into semantically meaningful partitions, an essential step in numerous applications \cite{chen2020deep,litjens2017survey} including diagnosis of several major cardiovascular diseases, such as dysplasia, cardiomyopathies, and pulmonary hypertension \cite{caudron2011diagnostic,caudron2012cardiac}.
Clinical data analysis can be tedious and time-consuming, with manual annotation of cardiac boundaries across different views and cycles. With the advent of deep learning, many advanced neural network-based algorithms have been proposed to automate cardiac image segmentation  \cite{baumgartner2018exploration,avendi2016combined,zheng20183,ringenberg2014fast}. However, the majority of these techniques only utilize the imaging modality as an input to the deep learning models, ignoring image-specific characteristics like acquisition parameters (scanner, vendor, field strength, number of frames, image quality), medical condition of the patients (disease, blood volume in ventricles, ejection fraction) and demographic specifications (sex, age).

\begin{figure}[t!]
\centering
\includegraphics[scale=0.37]{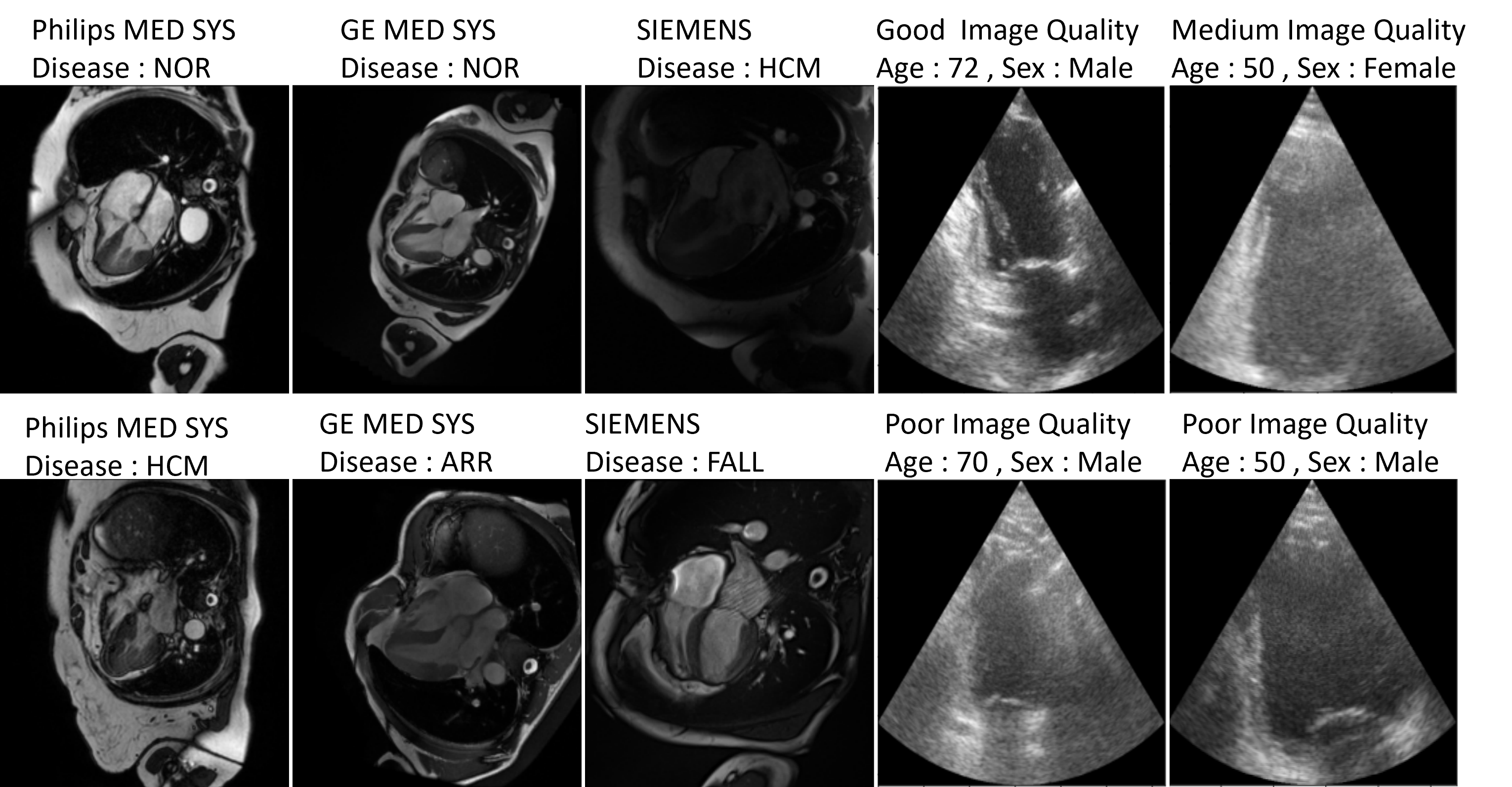}
\caption{Influence of metadata on image quality, appearance, and intensity patterns: Analysis of MRI (M\&Ms-2 dataset) and Ultrasound (CAMUS dataset) scans across different acquisition parameters and patient-specific information, i.e, disease, sex, and age.} 
\label{fig0}
\end{figure}

As shown in \figref{fig0}, acquisition parameters such as vendor and scanner and image-related characteristics, like disease, can affect image quality, appearance, and intensity patterns. For instance, images from Philips scanners in the dataset have higher intensity values than those from General Electric (GE) or Siemens. Also, images from patients without a disease (NOR) have a distinct intensity pattern when compared to images with underlying heart conditions, such as those affected by Hypertrophic Cardiomyopathy (HCM), Arrhythmogenic Cardiomyopathy (ARR), or Tetralogy of Fallot (FALL), for example, HCM may result in thicker ventricular walls, and ARR images show irregular heart shapes. 
% A good image quality indicates anatomical structures, high contrast, and fewer artifacts than medium and poor-quality images. 
In addition, the physiological aspects, such as age and sex, also contribute to the image characteristics; for example, older age patients may have poor image quality due to factors such as changes in tissue density and sex affects the heart size and position.
%ing in the scanning process and image interpretation. 
Incorporating these correlations between the images and metadata within training can aid a segmentation model in accurately identifying patterns within the imaging data, resulting in improved robustness and accuracy.
%Also, most medical image segmentation methods use a single encoder-decoder network to segment complex anatomies or two cascaded models for more accurate segmentation. The cardiac anatomies segmentation methods first need to locate the heart tissue from an entire cardiac image as the accusations scanner captures not only the heart but the surrounding areas as well, and then perform the segmentation.

Existing deep learning methods for medical image segmentation can be divided into two categories: (1) single-stage methods and (2) two-stage methods. For single-stage methods, the entire image is directly fed to the network \cite{ronneberger2015u,chen2021transunet}. For two-stage approaches, the search area is limited by localizing the organ(s) and further segmenting each class \cite{dogan2021two,zhang2021cascade}. %, such as cascaded networks. 
The single-stage methods are efficient regarding the end-to-end training and inference time. However, they may struggle to precisely segment the cardiac regions due to their complex anatomy,  motion, high variability in shape between individuals, and the challenges posed by different image quality and modalities. While two-stage methods, which involve a localization network (coarse segmentation or a regression network) followed by a detailed segmentation network, can achieve higher accuracy by focusing on the relevant areas and refining the segmentation, they increase computational complexity, requiring more resources and time for inference and training.

This paper addresses these challenges to enable accurate anatomical segmentation. To this end, we propose a super-to-sub segmentation compositional approach to localize and segment the heart simultaneously and a Cross-Modal Feature Integration (CMFI) module to incorporate the metadata associated with each cardiac image to handle the variability in image characteristics that result from different equipment, protocols, patient conditions and demographics depending upon the metadata availability. We validated the proposed method on two different imaging modalities, MRI and Ultrasound, using Multi-Disease, Multi-View, and Multi-Centre (M\&Ms 2) \cite{campello2021multi,martin2023deep} and Multi-structure Ultrasound Segmentation (CAMUS) dataset \cite{leclerc2019deep}, respectively.

%limited clinical application due to their restricted training and evaluation criteria; often, training data are from a single center, consisting of healthy subjects, with single-view scans, and similar imaging protocols are used.  Campello et al. \cite{campello2021multi} released a Multi-Disease, Multi-View, and Multi-Centre (M\&Ms-2)\footnote{https://www.ub.edu/mnms-2/} dataset to highlight challenges within a multi-center and scanner vendor setting. This paper addresses these challenges to enable accurate anatomical segmentation using our compositional segmentation approach to localize and segment the heart. Furthermore, our method incorporates the metadata associated with each MR image to handle the variability in image and characteristics that result from different equipment and protocols.

\section{Related work}
\label{sec:relatedwork}
UNet \cite{ronneberger2015u} pioneered an encoder-decoder model architecture for medical image segmentation. A number of methods have been proposed to further improve UNet \cite{diakogiannis2020resunet,isensee2021nnu,li2021right}. The `No New-Net UNet' (nnUNet) \cite{isensee2021nnu} is an extension to UNet with automatic hyperparameter configuration to target a range of medical image segmentation tasks. InfoTrans \cite{li2021right} utilized nnUNet for cardiac segmentation, where information transition was proposed to utilize the long-axis (LA) view to assist with the segmentation of a short-axis (SA) view. The predicted LA views were utilized to locate and crop the SA views. Tempera \cite{galazis2021tempera} proposed a Spatial Transformer Feature Pyramid-based Network that uses hybrid 2D/3D convolutions to segment the right ventricle (RV). A multi-view SA-LA model is proposed by Jabbar et al. \cite{jabbar2021multi} to segment the RV on the SA and LA cardiac MR images. Their method is trained and validated on 2D slices from MRI volumes, where the bottleneck layers of both views are coupled as input to the decoder.

Recent improvements shown by vision transformers \cite{dosovitskiy2020image} have inspired several medical image segmentation architectures \cite{chen2021transunet,gao2021utnet,ji2021multi,liu2022transfusion}. TransUNet \cite{chen2021transunet} improved the UNet architecture with self-attention within the encoder only. UTNet \cite{gao2021utnet} proposed an efficient self-attention mechanism with reduced computational complexity, incorporating self-attention in both the encoder and decoder. MCTrans \cite{ji2021multi} utilized multi-view inputs and performs intra- and inter-scale self-attention of different convolutional features. TransFusion \cite{liu2022transfusion} merged multi-view imaging information using a Divergent Fusion Attention (DiFA) to capture long-range correlations between unaligned data and a Multi-Scale Attention block for learning the global correspondence of multi-scale feature representations.

Our proposed compositional approach and metadata utilization strategy are inspired by several methods, including cascaded architectures, \cite{dogan2021two,xue2019cascaded}, FiLMed-UNet \cite{lemay2021benefits}, and SwiftFormer \cite{shaker2023swiftformer}. 
Similar to two-stage methods \cite{dogan2021two,xue2019cascaded}, firstly, the heart is localized using \emph{super-segmentation} decoder, and then \emph{sub-segmentation} decoder simultaneously segments the heart into LV, LA, RV, and MYO.
Lemay et al. proposed FiLMed-UNet \cite{lemay2021benefits}, where the authors used Feature-wise Linear Modulation (FiLM) layers to integrate metadata at different encoder-decoder stages of a UNet, leading to improved segmentation accuracy. However, our proposed approach learns metadata as an auxiliary task, using a classifier based on Multilayer Perceptron (MLP) \cite{haykin1998neural}. We also propose a novel way of integrating the metadata features into the segmentation network using the CMFI module.

Our work is also inspired by \cite{shaker2023swiftformer}, which proposes an efficient additive attention (E-2A) mechanism to reduce the quadratic computational complexity of self-attention to a linear element-wise multiplication. In our proposed CMFI module, we have utilized the E-2A to intermingle the image and metadata features.  More specifically, we perform cross-attention between the segmentation network features and metadata features to provide additional context about the image's content and improve the segmentation accuracy. Our contributions are as follows:

\begin{enumerate}[topsep=-0.5pt]
    \item We propose a novel compositional segmentation approach that simultaneously localizes the heart (super-segmentation) and segments the heart structures (sub-segmentation). 
    \item We propose a Cross-Modal Feature Integration (CMFI) module to utilize the image metadata, including acquisition parameters, medical condition, and demographic of the patient, to conditionally modulate the segmentation network. 
    \item Extensive quantitative and qualitative experimental comparisons demonstrate that our proposed method outperforms the existing state-of-the-art. We evaluate the proposed approaches on two different modalities, MRI and ultrasound, and show that our approach excels in these diverse domains. The consistent performance improvements observed in both modalities indicate that our method could yield similar accuracy enhancements in other domains and modalities.  
    %We also include multiple ablation studies to demonstrate the effectiveness of the proposed compositional segmentation and CMFI module-based conditioning approaches 

\end{enumerate}
\section{Methods}
\begin{figure*}[t!]
\centering
\includegraphics[scale=0.42]{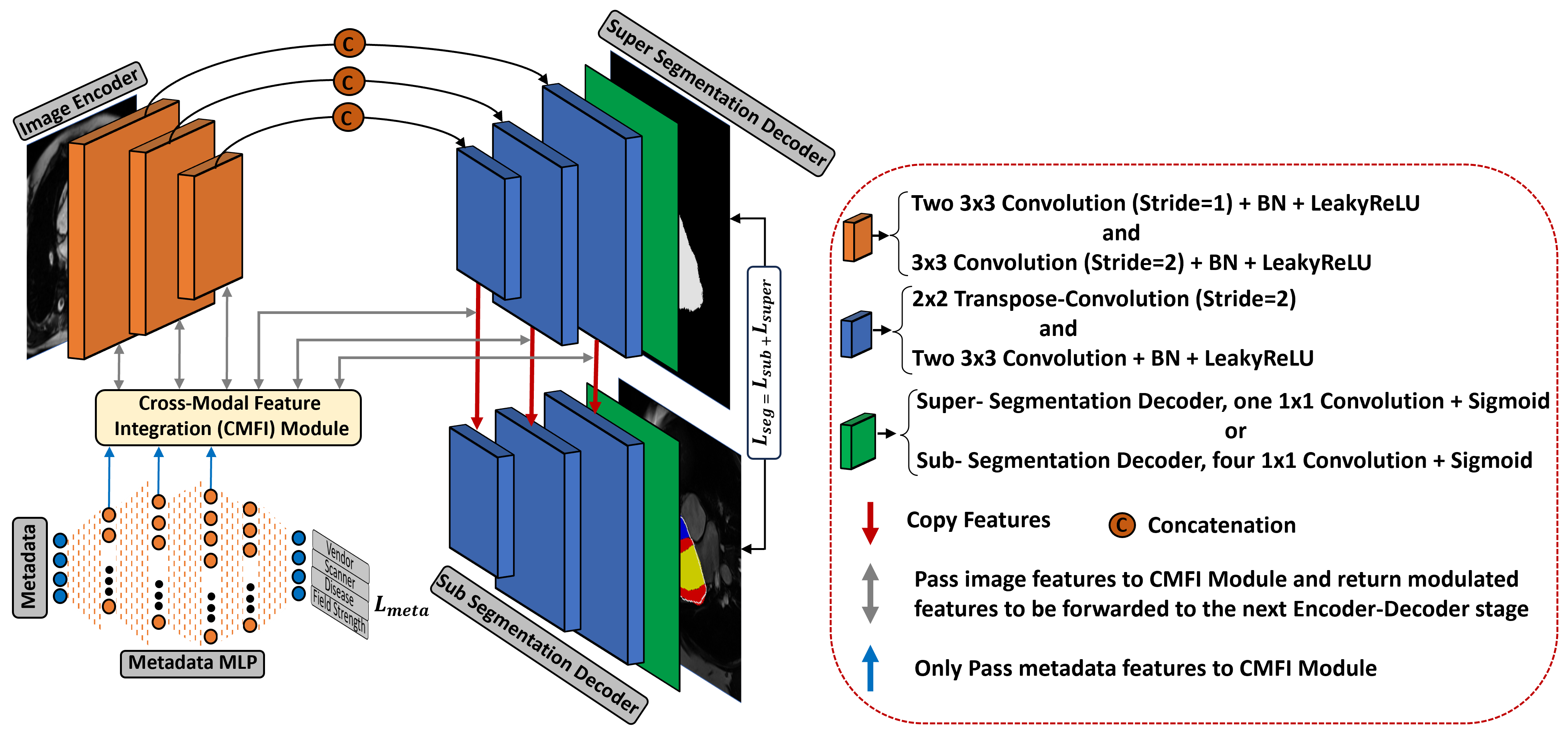}
\caption{Overview of the proposed pipeline. The network has five encoder-decoder stages; only three are shown here for simplicity. The image encoder extracts features from the image; the two decoders perform super and sub-segmentation. The metadata is learned via MLP, followed by the interaction of image and metadata features using the CMFI module.} 
\label{fig1}
\end{figure*}
\figref{fig1} shows the proposed multi-task compositional cardiac image segmentation network. The inputs to the model are both cardiac images and related metadata information. The image encoder extracts features from the image, and the metadata is learned via an MLP. Our proposed CMFI module conditions the segmentation network based on the metadata for a given image. Here, we emphasize that a medical image, such as a cardiac MRI, is influenced by characteristics of the imaging equipment employed for the acquisition of the image, such as manufacturer (vendor), scanner type, field strength, image quality, underlying anatomical characteristics (pathological conditions,  blood volume in ventricles) depicted in the image \cite{onofrey2019generalizable}, and patient's demographics. Hence, the image segmentation network is guided based on the hallmarks associated with the intensity images to enhance the segmentation performance.

In the proposed hierarchical decoder strategy, the super-segmentation decoder gets the modulated features and localizes the heart as a single region (super-segmentation). As a simultaneous step, the sub-segmentation decoder copies the decoder features from the super-segmentation and refines part-based segmentation further into different Regions of Interest (ROIs), such as LV, LA, RV, and MYO (sub-segmentation). Our compositional segmentation network is trained end-to-end simultaneously for both super and sub-segmentation and classification/regression tasks.

\subsection{Encoders Strategy}
\textbf{Image Encoder:} Convolutional layers are used to extract features from the image. Each block consists of two consecutive 3$\times$3 convolutions followed by batch normalization and LeakyReLU activation \cite{hendrycks2016gaussian}. The number of feature maps for the image encoder is increased to {32, 64, 128, 256, and 320} while down-sampling spatially by a factor 2$\times$ using a 3$\times$3 convolution with stride 2. 

\textbf{Metadata MLP:} For metadata, an MLP is implemented, shown in the bottom left of \figref{fig1}. The metadata is passed through a series of linear layers where the number of linear layers equals the number of encoder or decoder stages, and the number of neurons at each layer is equal to the number of feature maps at the respective encoder-decoder stage. Every linear layer is followed by 1D-batch normalization, LeakyReLU activation, and a dropout of 0.1. Finally, a linear layer transforms the high-dimensional feature space into a 128-dimensional representation and feeds it into the respective linear layer to classify or regress the available metadata entity.

\subsection{Encoding MetaData Into a Metadata Tensor}
The M\&Ms-2 dataset was captured using MRI machines from three vendors and nine scanners under two magnetic field strengths, 1.5 and 3 Tesla, to create a highly heterogeneous dataset that reflects the diversity seen in real-world clinical practice. The training set has five diseases and instances of normal cases.
Each metadata entity is mapped to a numerical representation; for example, vendors {Philips, Siemens, and GE} are mapped to numerical values (e.g., 1, 2, 3) using a predefined dictionary. A similar mapping procedure is adopted for scanner and disease categories. The field strength, i.e., (1.5 or 3 Tesla), is used directly.

The metadata extracted from the CAMUS dataset includes attributes such as end-systolic (ES) and end-diastolic (ED) frames, the total number of frames (NbFrame), patient sex and age, image quality, ejection fraction (EF), and frame rate. We normalized continuous metadata values by dividing them by a factor of 10 to scale input features appropriately. Categorical variables, such as sex and image quality, were mapped to numerical values; for example, sex:\{Male, Female\} mapped to {0,1} and image quality:\{Good, Medium, Poor\} to {0,1,2}.  All metadata encodings are released with our source code.
\subsection{Cross-Modal Feature Integration (CMFI) module} 
\begin{figure}[t!]
\centering
\includegraphics[scale=0.50]{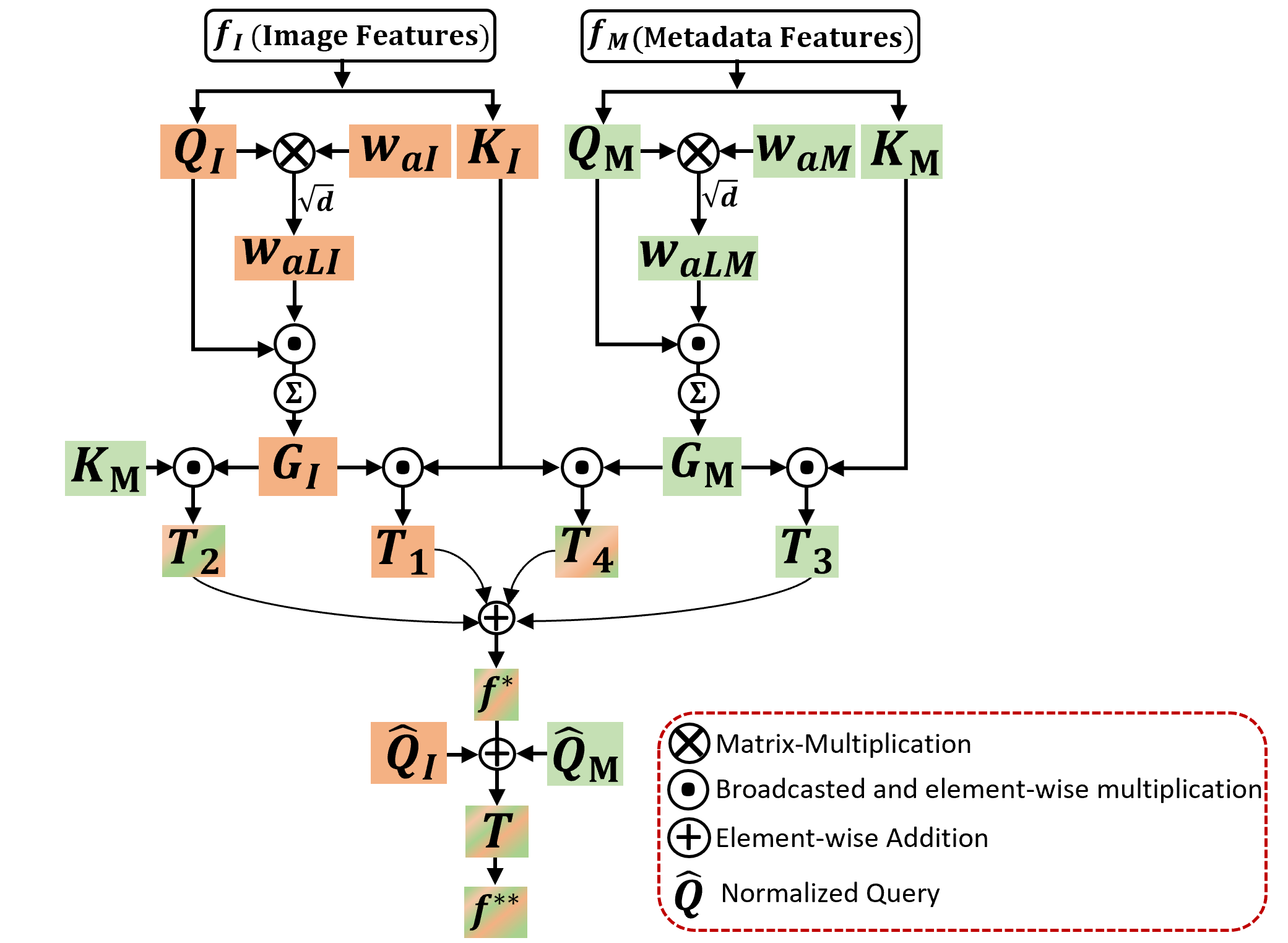}
\caption{The proposed CMFI Module. Each block's subscripts, $\boldsymbol{I}$ and $\boldsymbol{M}$, represent the image and metadata features, respectively.} 
\label{cmfi_fig}
\end{figure}

The proposed CMFI module, shown in \figref{cmfi_fig} requires the Query (${\mathbf{Q}}$) and Key (${\mathbf{K}}$) interaction and is used at all stages of the segmentation network. At each encoder-decoder stage, the metadata feature matrix \(({\mathbf{\textbf{f}}_M})\) of shape $\mathbb{R}^{B \times C}$ is expanded to the size of the image feature matrix \(({\mathbf{\textbf{f}}_I})\), resulting in identical dimensions of $\mathbb{R}^{B \times C \times H \times W}$, where ($B$: Batch size, $C$: Number of channels, $H$: Height, $W$: Width). Finally, the features from both modalities are reshaped to $\mathbb{R}^{B \times N \times C}$, such that,  $({\textbf{f}_{I}},{\textbf{f}_{M}}) \in \mathbb{R}^{B \times N \times C}$ , and $N = H \times W$.

Each feature matrix \({\mathbf{\textbf{f}}_M}\) and \({\mathbf{\textbf{f}}_I}\) undergoes a projection to generate corresponding ${\mathbf{Q}}$ and ${\mathbf{K}}$ matrices, i.e., $({\mathbf{Q}_I}$,${\mathbf{K}_I}$: for image features, and ${\mathbf{Q}_M}$, ${\mathbf{K}_M}$: for metadata features), such that, $({\textbf{Q}_{I}},{\textbf{K}_{I}},{\textbf{Q}_{M}},{\textbf{K}_{M}}) \in \mathbb{R}^{B \times N \times C}$

 The learnable attention weights for image modality $({\textbf{w}_{aLI}}\in \mathbb{R}^{B \times N \times 1})$ and metadata modality $({\textbf{w}_{aLM}}\in \mathbb{R}^{B \times N \times 1})$ are obtained by following steps.

 \begin{enumerate}
     \item Multiply the query matrix of the image modality (${\mathbf{Q}_I}$)  with its corresponding parameter vector 
     (${\textbf{w}_{aI}}\in \mathbb{R}^{C \times 1}$) to get $({\textbf{w}_{aLI}})$.
     \item  Multiply the query matrix of the metadata modality (${\mathbf{Q}_M}$)  with its corresponding parameter vector (${\textbf{w}_{aM}}\in \mathbb{R}^{C \times 1}$) to get $({\textbf{w}_{aLM}})$.
     \item Apply a scaling operation to these products.
 \end{enumerate}

Mathematically, these operations can be represented as:
\begin{equation}\label{eq1}
{\textbf{w}_{aLI}}  =  \frac{\mathbf{{Q}_I}  \textbf{w}_{aI}}{\sqrt{C}}      \quad\quad \textrm{and} \quad\quad  {\textbf{w}_{aLM}}  =  \frac{\mathbf{{Q}_M}  \textbf{w}_{aM}}{\sqrt{C}}  
\end{equation}
%where ${d}_{I}$, and ${d}_{M}$ are the dimensions of $\mathbf{Q}_{I}$, and $\mathbf{Q}_{M}$, respectively.
Following this, the (${\textbf{w}_{aLI}}$,${\textbf{w}_{aLM}}$) are multiplied with (${\mathbf{Q}_I}$,${\mathbf{Q}_M}$)  and summed along dimension $N$, to produce a single global attention query vector for image modality $({\textbf{G}_{I}}\in \mathbb{R}^{B \times C})$, and metadata modality $({\textbf{G}_{M}} \in \mathbb{R}^{B \times C})$:
\begin{equation}\label{eq1a}
{\mathbf{G}_{I}}  =  \sum_{n=1}^{N} (\textbf{w}_{aLI})_{N}  \odot \mathbf{Q}_{IN} 
\end{equation}
and,
\begin{equation}\label{eq10}
{\mathbf{G}_{M}}  =  \sum_{n=1}^{n} (\textbf{w}_{aLM})_{n}  \odot \mathbf{Q}_{Mn}
\end{equation}

where $\odot$ denotes element-wise multiplication, and global attention query vectors are expanded to the dimension of $\mathbb{R}^{B \times N \times C}$. The global and cross-global context for ${\mathbf{\textbf{f}}_I}$ is established through the interaction of ${\mathbf{G}_{M}}$ and ${\mathbf{G}_{I}}$ with both ${\mathbf{K}_I}$, and ${\mathbf{K}_M}$, followed by a linear transformation layer ($\fbseries\boldsymbol{T}_{j}$) for the $\fbseries j$th pair of interactivity, where $\fbseries j \in {1,2,3,4}$.
\setBoldness{0.5}%
\begin{equation}\label{eq1b}
\begin{split}
{\mathbf{f}^{*}_{I}} = & \, {\mathbf{T}_{1}}({\mathbf{G}_I} \odot {\mathbf{K}_I}) + {\mathbf{T}_{2}}({\mathbf{G}_I} \odot {\mathbf{K}_M}) + \\
& \, {\mathbf{T}_{3}}({\mathbf{G}_M} \odot {\mathbf{K}_M}) + {\mathbf{T}_{4}}({\mathbf{G}_M} \odot {\mathbf{K}_I})
\end{split}
\end{equation}
Finally, the normalized $\mathbf{{Q}_I}$, and $\mathbf{{Q}_M}$ are summed with $\textbf{f}^{*}_{I}$, and another linear transformation layer ($\fbseries\boldsymbol{T}$)\hspace{0.1em} is applied on the resultant vector to obtain the final modulated segmentation network features ${\mathbf{\textbf{f}}^{ **}_{I}}$ which is passed to the next encoder-decoder stage: 
\begin{equation}\label{eq1c}
{\mathbf{\textbf{f}}^{ **}_{I}}  = {\fbseries\boldsymbol{T}}\Bigl({\mathbf{\textbf{f}}^{ *}_{I}} + \frac{\mathbf{Q}_I}{\| \mathbf{{Q}_I} \|_2} + \frac{\mathbf{Q}_M}{\| \mathbf{{Q}_M}\|_2} \Bigl)
\end{equation}
where, $\| \cdot \|_2$ denotes the Euclidean norm.

\subsection{Hierarchical Decoder Strategy}
Our proposed hierarchical decoder strategy comprises two decoders: super and sub-segmentation decoders. Each decoder upsamples the features using 2$\times$2 transpose convolutions, followed by two 3$\times$3 convolutions, batch normalization, and LeakyReLU activation. 
The super-segmentation decoder utilizes skip connections from the encoder to segment all three ROI classes as a single binary segmentation map. 
The sub-segmentation decoder gets the features from the super-segmentation decoder and further segments the binary segmentation features into multiple classes. This feature-copying process from the super-segmentation decoder (shown by the red downward arrow in \figref{fig1}) makes part-based segmentation of ROIs more efficient. Instead of directly looking at the entire image and finding the relevant features, our approach implicitly confines the search area based on super-segmentation. Empirical tests showed that adding skip connections to the sub-segmentation decoder did not improve its overall accuracy and added extra learnable weights. Therefore, the sub-segmentation decoder does not utilize skip connections.

The proposed multi-task network is trained end-to-end using the composite loss function below.
%$\mathcal{L}_{total}$ which is defined as 
%$\mathcal{L}_{total}=  \alpha \mathcal{L}_{cls} + (1-\alpha) \mathcal{L}_{seg}$,
\begin{equation}\label{eq_loss1}
\mathcal{L}_{total}=   \alpha \mathcal{L}_{seg} + (1-\alpha) \mathcal{L}_{meta},
\end{equation}
where $\mathcal{L}_{seg}$ is the segmentation loss associated with the proposed compositional approach, and  $\mathcal{L}_{meta}$ is the loss to learn the metadata entities, described below. The segmentation loss $\mathcal{L}_{seg}$ is composed of the sub-segmentation loss ($\mathcal{L}_{sub}$) and the super-segmentation loss ($\mathcal{L}_{super}$).
\begin{equation}\label{eq_loss2}
\mathcal{L}_{seg}=  \mathcal{L}_{sub} + \mathcal{L}_{super}.
\end{equation}
\begin{table*}[t!]
\centering
\caption{Comparison of the results obtained from different methods using LA views of M\&Ms-2 dataset using a five-fold cross-validation split. Methods indicated with a $*$ use multi-view inputs. The best results are shown in \textbf{Bold}.}\label{tab1}
        \resizebox{0.7\textwidth}{!}{
        \begin{tabular}{|c|c|c|c|c|c|c|c|c|}
            \hline
             \multirow{2}{*}{Methods} &  \multicolumn{4}{|c|}{Dice Score ($\%$) $\uparrow$} & \multicolumn{4}{|c|}{HD (mm) $\downarrow$}\\
             \cline{2-9}
              & LV & RV & Myo & Avg & LV & RV & Myo & Avg\\ 
            \hline
            UNet\cite{ronneberger2015u}    & 87.26 & 88.20 & 79.96 & 85.14 & 13.04 & 8.76 & 12.24 & 11.35\\
            ResUNet\cite{diakogiannis2020resunet}   & 87.61 & 88.41 & 80.12 & 85.38 & 12.72 & 8.39 & 11.28 & 10.80\\
            InfoTrans*\cite{li2021right} &  88.21 & 89.11 & 80.55 & 85.96 & 12.47 & 7.23 & 10.21 & 9.97\\
            TransUNet\cite{chen2021transunet}  & 87.91 & 88.23 & 79.05 & 85.06 & 12.02 & 8.14 & 11.21 & 10.46\\
            MCTrans\cite{ji2021multi}   & 88.42 & 88.19 & 79.47 & 85.36 & 11.78 & 7.65 & 10.76 & 10.06\\
            MCTrans*\cite{ji2021multi}  &  88.81 & 88.61 & 79.94 & 85.79 & 11.52 & 7.02 & 10.07 & 9.54\\
            UTNet\cite{gao2021utnet} &  86.93 & 89.07 & 80.48 & 85.49 & 11.47 & 6.35 & 10.02 & 9.28\\
            UTNet*\cite{gao2021utnet} &  87.36 & {90.42} & 81.02 & 86.27 & 11.13 & {5.91} & {9.81} & {8.95}\\
            TransFusion*\cite{liu2022transfusion}   & {89.78} & {91.52} & {81.79} & {87.70} & {10.25} & {5.12} & {8.69} & {8.02}\\ 
            UNETR\cite{hatamizadeh2022unetr}   & {91.33} & {85.71} & {80.85} & {85.96} & {10.08} & {11.28} & {6.07} & {9.14}\\
            SWIN-UNETR\cite{hatamizadeh2021swin}  & {92.21} & {86.93} & {81.77} & {86.96} & {8.84} & {9.73} & {5.60} & {8.05}\\
            nnUNet*\cite{isensee2021nnu}   & {94.08} & {90.70} & {86.41} & {90.39} & {5.91} & {6.61} & {5.98} & {6.16}\\ 
             Two-Stage\cite{xue2019cascaded}   & {94.20} & {89.15} & {85.51} & {89.62} & {4.40} & {6.29} & {3.90} & {4.86}\\  \hline
             Proposed (WO/ Super-Seg.) & {90.57} & {88.05} & {82.19} & {86.93} & {6.94} &  {8.41} & {5.22} & {6.85}\\
              Proposed (WO/ CMFI) & {94.48} & {90.29} & {85.31} & {90.02} & {3.55} &  {5.19} & {\textbf{2.64}} & {3.80}\\
             \textbf{Proposed W/(Super-Seg.+CMFI)} & \textbf{95.63} &  \textbf{91.93} &  \textbf{87.61} &  \textbf{91.72} &  \textbf{3.16} &  \textbf{4.62} & 2.95 &  \textbf{3.57}\\
            \hline
        \end{tabular}}
\end{table*} 
For both datasets, the segmentation losses, i.e, $\mathcal{L}_{sub}$ and  $\mathcal{L}_{super}$ are the Dice losses.
For the M\&Ms-2 dataset, the $\mathcal{L}_{meta}$ loss is a cross-entropy loss for classification employed for the MLP network. For the CAMUS dataset, the $\mathcal{L}_{meta}$ is the sum of cross-entropy and L1-loss depending on the nature of metadata, i.e., for continuous variables, L1-loss and for categorical, the cross-entropy loss.

Based on empirical experiments, $\alpha$ is the balancing factor in equation (6) between the two loss terms and is set to $\alpha=0.7$. Given the simpler nature of the classification task, which serves as an auxiliary component compared to segmentation, a higher weight is assigned to the segmentation task in the overall loss formulation.

Here, we emphasize that our proposed compositional approaches (super and sub-segmentation) and CMFI module for incorporating the metadata methods are general purposes and can be applied to other segmentation networks as well. In this paper, we have used the UNet architecture as a baseline with some modifications, including strided convolutions with stride = 2 to reduce the in-plane spatial dimensions of features in the encoder compared to 2 $\times$ 2 max-pooling operations of UNet, transpose convolutions in the decoder to upsample the features compared to UNet's bilinear interpolation method and replacing the ReLU activation with LeakyReLU. 

\section{Experimental Validation}
In this section, we provide details of the datasets, implementation and our experimental validation results showing our approach's superior performance as compared to the state-of-the-art.

\subsection{Datasets Description}
We utilize the following two datasets for the experimental validation of our proposed methods:

\noindent \textbf{M\&Ms-2 data}: This data comes from a challenge cohort hosted by MICCAI 2021 \cite{campello2021multi,martin2023deep}, consisting of RV blood pool segmentation across cardiac MRI imaging of SA and LA views. Segmentation labels are provided for three ROIs: (i)  LV blood pools, (ii) RV blood pools, and (ii) LV myocardium (LV-MYO). In our work, we conduct LA view segmentation experiments from M\&Ms-2 data. Following \cite{liu2022transfusion}, we randomly shuffled the 160 training samples and evaluated all models using a 5-fold cross-validation split. 

\noindent \textbf{CAMUS data}: This dataset \cite{leclerc2019deep} provides 2D echocardiographic images of two and four-chamber views for 500 patients. The CAMUS provides manual labels for the left ventricle endocardium (LV), the myocardium epicardium  (MYO), and the left atrium (LA). In the proposed study, we have utilized the two-chamber views from 500 patients in a 5-fold cross-validation split.
\subsection{Implementation Details}
The proposed framework is implemented using PyTorch and an NVidia A100 GPU with 40GB RAM. All models are trained using Adam optimizer \cite{kingma2014adam} for 500 epochs, learning rate = $1e^{-4}$ and batch size 16. The images are resampled to an in-plane resolution of 1.25$\times$1.25 ${mm^2}$ for M\&Ms-2 and 1$\times$1 ${mm^2}$ for CAMUS data. Each image is normalized by its mean and standard deviation.  Various geometric and intensity data augmentation strategies are utilized, including rotation, shift, scaling, elastic deformation, Gaussian noise, Gaussian blur, and random bias field.

\begin{figure*}[t!]
\centering
\includegraphics[scale=0.40]{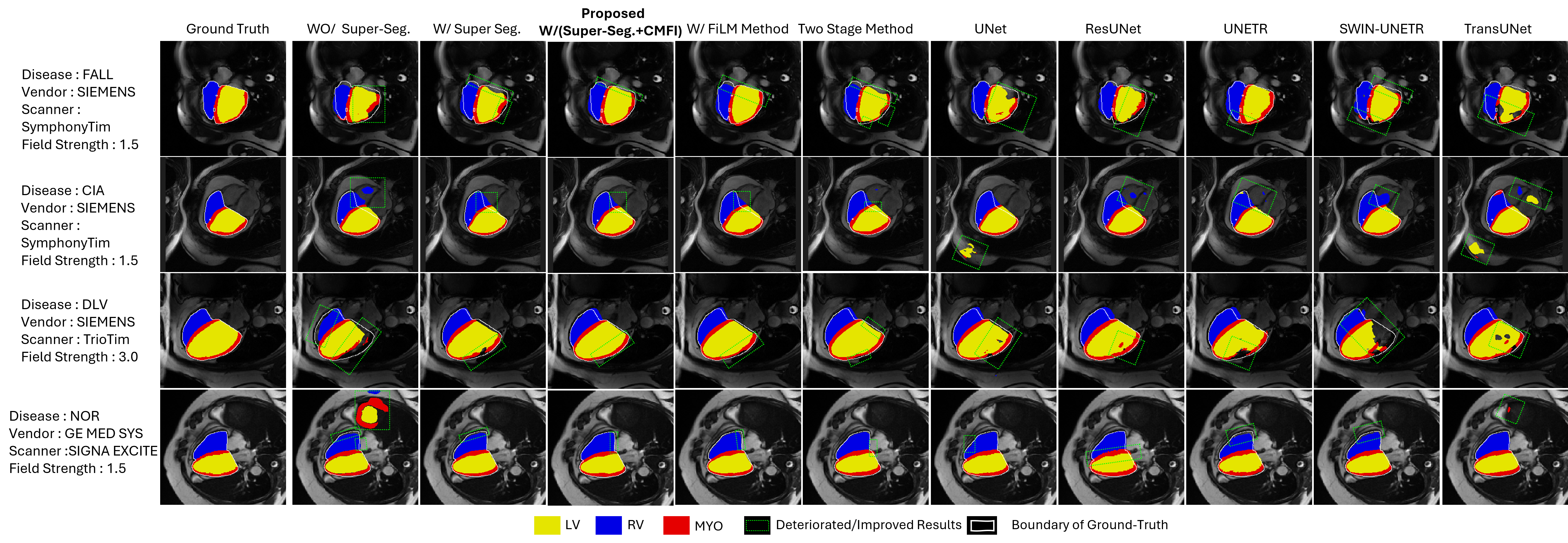}
\caption{Visual comparison of our compositional approach with (W/) and without (WO/) the super-segmentation, metadata utilization strategy using FiLM\cite{lemay2021benefits}, proposed CMFI module, and other comparative networks using M\&Ms-2 dataset. Please zoom in for details.} 
\label{VisaulResults}
\end{figure*}

\subsection{Experimental Validation with M\&Ms-2 Dataset}
Our experimental validation results with the M\&Ms-2 dataset, shown in \tabref{tab1}, compare our proposed method with existing state-of-the-art methods, including UNet-based, transformer-based, and two-stage segmentation approaches. The proposed hierarchical segmentation method utilizes UNet and has a few extra trainable parameters compared to UNet. Figure 1 in the supplementary material depicts further insights into the performance and number of parameter methods.
%hese extra parameters result from the Metadata MLP and sub-segmentation decoder. 
%Moreover, we have utilized the stride convolutional layers instead of the max-pooling operation of UNet \cite{springenberg2014striving} in the proposed without super-segmentation (WO/ Super-Seg.) method. 
%\figref{modelSizeVsAccuracy} depicts the proposed method's computational complexity compared to the existing methods. With both proposed strategies, compositional approach, and CMFI-Module, our method achieved higher Dice scores than others by striking a good balance between parameters and performance.
Our proposed method achieves state-of-the-art accuracy with the highest Dice and lowest Hausdorff Distance (HD) scores across all segmentation classes. Furthermore, \figref{VisaulResults} shows qualitative results, showing the proposed compositional approach enables accurate delineation of segmentation while reducing false positives outside the heart region. This improved accuracy is mainly due to two aspects of our proposed: (i) hierarchical decoder where super-segmentation decoder, which confines the segmentation to the heart region and helps the sub-segmentation decoder by passing only relevant features, and (ii) incorporating additional metadata using the CMFI module that provides additional context resulting in specialized and accurate segmentation. We note that the accuracy of our hierarchical decoder is closer to a two-stage method; however, in contrast to the sequential execution of two-stage methods, where the first network localizes followed by segmentation, our compositional approach can simultaneously locate and segment heart and internal structures. 

% The improved HD score of the proposed compositional approach demonstrates its localization capability by producing more accurate boundary delineation. The qualitative results are shown in \figref{VisaulResults}; the proposed compositional approach can segment the boundaries accurately and reduce the false positives outside the heart region (outliers or floating predictions), the same as a two-stage method. However, compared to two-stage methods, where the first network performs localization, followed by the segmentation network, our compositional approach can simultaneously locate and segment the heart.
% This improved performance is because of the super-segmentation decoder, which confines the segmentation to the heart regions and helps the sub-segmentation decoder by passing only the relevant features, and the CMFI module, which integrates metadata into the segmentation network by providing additional context and generating more specialized and accurate segmentation.

\begin{table*}[ht!]
\centering
\caption{Quantitative results on five-fold cross-validation split of CAMUS data comparing performance with (W/) and without (WO/) super-segmentation decoder, metadata W/  CMFI module, and other comparative segmentation networks.}\label{ultra}
\resizebox{1.0\textwidth}{!}{
\begin{tabular}{|c|c|c|c|c||c|c|c|c|c|}
\hline
    Methods & LV-Dice  &   Myo-Dice & LA-Dice & Avg-Dice & LV-HD  & Myo-HD   & LA-HD  & Avg-HD \\  
    \hline
     UNet\cite{ronneberger2015u} &  91.52 & 84.70 & 86.44 & 87.55 & 19.78 & 21.22 &  32.02  & 24.34 \\
     ResUNet\cite{diakogiannis2020resunet} &  92.40 & 86.59 & 86.79 & 88.59 & 17.77 & 19.35 &  25.58  & 20.90 \\
     UNETR\cite{hatamizadeh2022unetr} &  91.64 & 84.81 & 86.93 & 87.79 & 16.06 & 17.68 &  22.44  & 18.72 \\
     TransUNet\cite{chen2021transunet}  &  88.58 & 80.79 & 81.65 &  83.67 & 31.54 & 42.18 & 34.89 & 36.20 \\
     SWIN-UNet\cite{cao2022swin} &  92.06 & 85.64 & 87.47 & 88.38 & 14.94 & 16.40 & 18.78 &  16.70  \\
    SWIN-UNETR\cite{hatamizadeh2021swin} &  92.60 & 86.55 & 87.03 & 88.72 & 15.98 & 16.80 & 19.95 & 17.57 \\
    \hline
     Proposed (WO/ Super-Seg.) &   92.29 & 86.06 & 88.31 & 88.88 & 15.06 & 17.38 &  22.18  & 18.20 \\
     
        Proposed (WO/ CMFI)  & {93.44} & {87.58} & {89.08} & {90.03} & {13.44} &  {15.52} & {18.19} & {15.71}\\ 
        
         Proposed W/(Super-Seg.+CMFI)  & \textbf{93.48} & \textbf{88.70} & \textbf{89.90} & \textbf{90.69} & \textbf{12.17} &  \textbf{15.01} & \textbf{17.89} & \textbf{15.02}\\ 
         \hline
    \end{tabular}
    }
\end{table*}
\begin{figure*}[h!]
\centering
\includegraphics[scale=0.36]{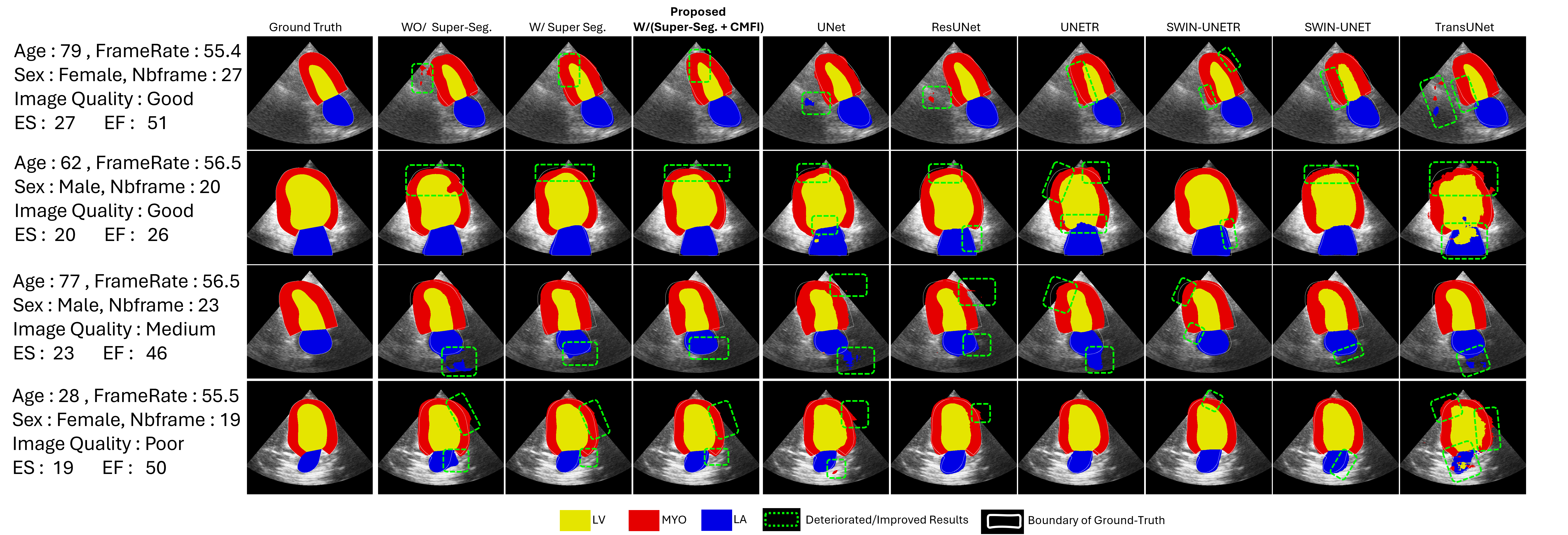}
\caption{Qualitative comparison of proposed compositional approach with (W/) and without (WO/) the super-segmentation and metadata utilization strategy using CMFI module using CAMUS data. Please zoom in for details.} 
\label{ultra_viz}
\end{figure*}

Strategically incorporating metadata alongside intensity images enhances the accuracy and reliability of the results. Metadata is motivated by its capacity to offer contextual information that image data alone cannot provide. For example, as shown in \figref{VisaulResults}, the first two rows come with FALL and Inter-atrial communication (CIA) diseases, which primarily impacts the RV outflow tract obstruction and volume overload, respectively. In the W/ CMFI Module column, where we supplied the model with this disease-specific information, it can prioritize the segmentation of the RV, ultimately leading to improved delineation and accuracy of the RV. Similarly, in the third row, metadata about the Dilated Left Ventricle (DLV) guides the model in accurately segmenting the LV and the MYO, acknowledging the expected dilation and MYO wall thinning associated with this condition. Furthermore, metadata such as vendor, scanner, and field strength have a crucial role in adapting the model to the underlying variations in image characteristics. The CMFI module is designed to merge metadata with images effectively, which enables the model to make more informed predictions by leveraging the additional data. This leads to improved segmentation performance.
\subsection{Experimental Validation with CAMUS Dataset}
Our experiments on the CAMUS data \cite{leclerc2019deep} show the proposed methods' ability to work effectively in different modalities, where our proposed method achieves similar accuracy improvements as experiments with the M\&Ms-2 dataset.
\tabref{ultra} shows the quantitative results for CAMUS dataset, where the proposed super-segmentation decoder (without CMFI) yields notable improvements across all metrics for all three ROI. This is mainly due to better heart region localization as well as improved segmentation boundary delineation, 
as demonstrated in \figref{ultra_viz} (W/ vs WO/ super segmentation). The performance metrics improve further when the super-segmentation decoder and CMFI module are employed together, as listed in the last row of \tabref{ultra}. With the CMFI module and super segmentation decoder, the segmentation boundaries are further refined and accurately delineated, as shown in the fourth column of \figref{ultra_viz}. The CMFI module maintains performance with variable image quality across varying ages, genders and various settings for functional aspects of the cardiac images.
\begin{table*}[t!]
\centering
\caption{Ablation studies comparing performance W/ and WO/ super-segmentation decoder and metadata with FiLM \cite{lemay2021benefits} or CMFI using a five-fold cross-validation split of the M\&Ms-2 dataset. Our proposed method with the super-segmentation and CMFI is in the bottom row.}
\resizebox{0.9\textwidth}{!}{
\begin{tabular}{|c|c|c|c|c|c|c||c|c|c|c|c|}
\hline
    Super Segmentation & FiLM & CMFI & LV-Dice  & RV-Dice & Myo-Dice  & Avg-Dice & LV-HD  & RV-HD   & Myo-HD & Avg-HD \\  
    \hline
     & & &  90.57 & 88.05 & 82.19 & 86.93 & 6.94 &  8.41 &  5.22  & 6.85 \\
     \hline
        \usym{1F5F8} & &  & {94.48} & {90.29} & {85.31} & {90.02} & {3.55} &  {5.19} & {2.64} & {3.80}\\ 
        \hline
        \usym{1F5F8} & \usym{1F5F8} &  & {95.10} & {91.10} & {86.04} & {90.74} & {3.98} &  5.18 & \textbf{2.15} & 3.77 \\  
        \hline
        \usym{1F5F8} &  & \usym{1F5F8} &   \textbf{95.63} & \textbf{91.93} & \textbf{87.61} & \textbf{91.72} & \textbf{3.16} & \textbf{4.62} & 2.95 & \textbf{3.57}\\  
        \hline
    \end{tabular}
    }
\label{tab:ablation}
\end{table*}

\begin{table*}[!t]
\centering
\caption{Performance comparison WO/ and W/ disease inclusion in metadata. The experiments are conducted on a five-fold cross-validation split of the M\&Ms-2 dataset.}\label{ab_3}
\resizebox{1.0\textwidth}{!}{
\begin{tabular}{|c|c|c|c|c|c|c|c|c|}
    \hline
   &  \multicolumn{4}{|c|}{\textbf{WO/ inclusion of disease}} & \multicolumn{4}{|c|}{\textbf{W/ inclusion of disease}}\\
    \hline
     Methods &  LA-Dice & RV-Dice &  Myo-Dice & Avg-Dice&  LA-Dice & RV-Dice &  Myo-Dice & Avg-Dice  \\
    \hline
    {Metadata Utilization W/ FiLM Method }  & {94.87} & {90.91} & {85.45} & {90.41} & \textbf{95.10} & \textbf{91.10} & \textbf{86.04} & \textbf{90.74} \\
    \hline
    {Metadata Utilization W/ CMFI module }  & {95.61} & {91.01} & {86.72} & {91.11} & \textbf{95.63} & \textbf{91.93} & \textbf{87.61} & \textbf{91.72} \\
    \hline
\end{tabular}
}
\end{table*}
Despite the good image quality in the first two rows of \figref{ultra_viz}, the method without metadata (third column)  results in extended LA boundaries and unclear separation between the MYO and LV, similarly, in rows three and four, where the image quality drops and results in blurred structures, the without metadata method further extends the MYO and LA due to unclear boundaries between the structures. However, when metadata is incorporated (fourth column), the model better understands the anatomical structures, leading to more precise boundaries and more explicit segmentation. Here, we speculate that the metadata, such as image quality, age, and sex, informs the model about the expected variations in size, motion, and function of the heart, making it more robust. For example, older individuals show thicker heart walls, reduced elasticity, and more tissue heterogeneity compared to younger. Also, males typically have larger heart chambers and thicker myocardial walls than females. The experimental settings can also help the network understand the patterns in the imaging data. For example, ES reflects the blood volume in the ventricle at the end of the contraction, and a higher value will indicate the larger size of LV, guiding the segmentation model to adjust the boundaries accordingly. 

%This demonstrates that combining patient-specific metadata with imaging data significantly improves the accuracy of heart structure segmentation.
%This adaptability suggests the module effectively integrates image quality metadata to enhance segmentation accuracy.
%The progressive improvements in Dice and HD scores affirm the efficacy of the proposed approaches in improving the segmentation performance using the CAMUS dataset.

\section{Ablation Studies}
The proposed compositional approach's effectiveness and metadata utilization are evaluated through the following ablation studies using the M\&Ms-2 dataset.

\noindent\textbf{Ablation without using the super-segmentation decoder.} In this ablation, we compare the effectiveness of our proposed method that utilizes a super-segmentation step to help localize the heart, shown in the first two rows of \tabref{tab:ablation}.
We report a Dice score ($>=$95\%) and $HD_{(mm)}$$<=$1.0 for super-segmentation (Not shown in tables as it's a pseudo-Dice score of three combined regions). The super-segmentation decoder improves the average Dice score of LV, RV, and MYO by 3.09\%  while reducing the average HD score by 3.05.

\figref{VisaulResults} visualizes how the super-segmentation decoder improves the segmentation accuracy of different regions of interest. By localizing the heart, we confined the search area for the sub-segmentation decoder (identifying an overall heart topology) to segment the LV, RV, and MYO. We note that the super-segmentation enables our proposed method to localize the relevant region of interest, resulting in improved segmentation accuracy.

\noindent\textbf{Ablation without using the CMFI module.} In this experiment, we show the effectiveness of the CMFI module to condition segmentation on the metadata information. We also compare against FiLM \cite{lemay2021benefits} and \tabref{tab:ablation} shows the accuracy comparison where we note that the network accuracy is improved by utilizing either of the approaches. However, the proposed CMFI module outperforms the FiLM method for all metrics with a 1.7\%  average improvement in the Dice score compared to 0.72\% of the FiLM method. This advocates that the proposed CMFI module provides a better way of leveraging the metadata through a non-linear attention mechanism and integrates metadata more nuancedly.

\noindent\textbf{Ablation without using the disease in metadata.} This ablation provides clinical justification and how the proposed approach can help to integrate clinical knowledge into the segmentation process. \tabref{ab_3} showcases the results W/ and WO/ using the disease information. The overall accuracy is improved by incorporating the disease into the segmentation network. This demonstrates that deep learning models can leverage clinical knowledge to guide the segmentation, especially where the goal is to prioritize the disease cases.

\noindent\textbf{Ablation to handle unavailability of metadata.} If \emph{no metadata is available}, we can use our method without CMFI, which still performs well compared to the existing methods, shown in the last two rows of \tabref{tab1}. If \emph{some metadata entity is missing during training}, e.g., disease, we can still use it by excluding the particular information and utilizing the rest of metadata, shown in \tabref{ab_3}. If \emph{some metadata entity is missing during inference}, e.g., vendor, we can run the model for each vendor Philips, Siemens, and GE used to train the model and then average the results to enhance the robustness by leveraging ensemble learning. 
%\begin{figure}[t!]
%\centering
%\includegraphics[scale=0.26]{figures/param.png}
%\caption{Model Size vs Average Dice score comparison of the proposed compositional and CMFI Module approach against different methods.} 
%\label{modelSizeVsAccuracy}
%\end{figure}
\section{Conclusion}
We proposed a compositional approach that simultaneously localizes the heart in cardiac images using a super-segmentation decoder and does part-based segmentation of different regions of interest through a sub-segmentation decoder. To leverage the image-specific metadata,
we also propose a CMFI module to integrate metadata into the segmentation network, guiding it with patterns associated with intensity images to improve performance.
%to guide the segmentation network based on the patterns associated with the intensity images, 
Extensive ablation studies indicate the efficacy of each proposed approach and compare it against existing state-of-the-art methods. The experiments are performed on two different modalities, MRI and ultrasound, using M\&Ms-2 and CAMUS datasets, respectively, to show that our approach works well in different domains thanks to the proposed compositional approach and CMFI module.

\newpage

%%%%%%%%% REFERENCES
{\small
\bibliographystyle{ieee_fullname}
\bibliography{egbib}
}

\end{document}